\documentclass[iop,apj]{emulateapj}
\slugcomment{To appear in the Astrophysical Journal}

\newcommand{\Msun}{\mbox{$M_{\odot}$}}
\newcommand{\Lsun}{\mbox{$L_{\odot}$}}
\newcommand{\Mearth}{\mbox{$M_{\oplus}$}}

\begin{document}
   \title{880 $\mu$m Imaging of a Transitional Disk in Upper Scorpius: Holdover from the Era of Giant Planet Formation?} 

\author{Geoffrey S. Mathews \altaffilmark{1}, Jonathan P. Williams \altaffilmark{1} , and Francois M\'enard \altaffilmark{2} 
}


\altaffiltext{1}{Institute for Astronomy (IfA), University of Hawaii,
              2680 Woodlawn Dr., Honolulu, HI 96822}
\altaffiltext{2}{Universit\'e Joseph-Fourier Ð Grenoble 1/CNRS, Laboratoire dÕAstrophysique de Grenoble (LAOG) UMR 5571, BP 53, 38041 Grenoble Cedex 09, France}
              \email{gmathews@ifa.hawaii.edu}
 
  \begin{abstract}

We present 880 $\mu$m images of the transition disk around the star [PZ99] J160421.7-213028, a solar-mass star in the nearby Upper Scorpius association.  With a resolution down to 0\farcs34, we resolve the inner hole in this disk, and via model fitting to the visibilities and spectral energy distribution we determine both the structure of the outer region and the presence of sparse dust within the cavity.  The disk contains $\sim$0.1 $\,M_{\rm Jup}$ ~of mm-emitting grains, with an inner disk edge of about 70 AU.  The inner cavity contains a small amount of dust with a depleted surface density in a region extending from about 20 -- 70 AU.  Taking into account prior observations indicating little to no stellar accretion, the lack of a binary companion, and the presence of dust near $\sim0.1$ AU, we determine that the most likely mechanism for the formation of this inner hole is the presence of one or more giant planets.  

  \end{abstract}


   \maketitle


\newpage

\section{Introduction}
\label{sec:intro}

It has been long suspected that planets form in disks of circumstellar material, with the hypothesis dating to \cite{Kant:1755}.  It was nearly two hundred fifty years, however, before tools were developed to allow the direct examination of this hypothesis.  Large scale studies of thermal emission from small dust grains in circumstellar disks were enabled by the Infrared Astronomical Satellite \citep[e.g.][]{1989AJ.....97.1451S}, while the development of millimeter observatories allowed the direct measurement of dust mass \citep{Beckwith:1990}.  The Spitzer space telescope allowed for statistically complete censuses of nearby star forming regions (e.g. c2d, FEPS, Gould's Belt Survey), while ground based interferometers such as the Smithsonian Sub-Millimeter Array (SMA) and Plateau de Bure Interferometer have allowed for imaging of the dust and gas components of the brightest disks \citep[e.g.][]{Andrews:2007,2009ApJ...701..698S}.  

Recent decades have also seen the discovery of hundreds of extrasolar planets.  Ground-based radial velocity and transit searches have determined that $\sim$10--20\% of stars host Jupiter mass and larger planets \citep{Johnson:2010}.  In the past year, the Kepler space telescope has revealed a similar fraction of stars hosting planets down to Neptune and Earth sizes \cite{Borucki:2011a,2011arXiv1103.2541H}.  Planet formation appears to be a common process, though only in the past year has a planet been observed in a young disk, confirming the connection between disks and planet formation \citep{2011arXiv1110.3808K}.  The formation timescale remains uncertain, however. 

A likely sequence is that the disk sees the growth of dust grains, which aggregate to form planetesimals and, over several millions of years, planets.  In the core accretion model, giant planets form in cases where the disk still has sufficient gas mass for large, $\sim$10 \Mearth, planetesimals to experience gas accretion, and then runaway growth, a process thought to occur in disks 3 -- 10 Myr old \citep{Pollack:1996}.  Whereas there are a handful of 10 Myr old massive disks, the median lifetime for infrared excess from disks is $\sim$3 Myr \citep{Williams:2011}.  However, the recent millimeter surveys of \cite{Lee:2011} and \cite{Mathews:2011} have found only a handful of stars in the IC 348 and Upper Scorpius regions hosting massive disks.  Most disks are already well advanced in the process of evolution from gas and dust rich primordial disks to sparse dust debris disks.  The paucity of massive disks in these regions compared with the relatively high frequency of exoplanets suggests that planet formation must be almost complete by the 3 to 5 Myr age of these regions.  

In addition to placing constraints on the timescale of formation, observations of disks may provide clues to systems where we have the best chance to directly observe planet formation in action.  A giant planet is expected to tidally open a gap in the disk \citep{Artymowicz:1994}, leading to inner disk dust depletion.  Such depletion is observable as a reduction in near-IR emission 
or a hole in emission which could be imaged with high resolution (sub)mm observations.
~Near to mid-infrared photometry and spectroscopy with Spitzer has shown that less than $\sim$10\% of stars with disks show some degree of inner disk depletion yet still retain massive outer disks \citep[e.g. ][]{2010ApJ...712..925C}.  To date, about a dozen such transitional disks around single stars have had their inner disk gaps verified by sub-millimeter imaging \citep{2007ApJ...664..536H, 2008ApJ...675L.109B, Andrews:2011c}.  

Here, we present spatially resolved 880 $\mu$m continuum and CO J=3$\rightarrow$2 images showing a transitional disk around a star in the nearby \object{Upper Scorpius} association.  At an estimated age of 5 \citep{Preibisch:2002}, Upper Scorpius lies in the mid-range of the theoretical timescale for giant planet formation, and at a distance of 145 pc \citep{de-Zeeuw:1999}, it is one of the closest regions of recent star formation.  Recently, a somewhat older age of $\sim$12 Myr has been suggested by \cite{Pecaut:2011}, which would place the association at the late-range of theoretical planet formation time-scales.

\label{sec:J1604}

The star \object{[PZ99] J160421.7-213028} (aka \object{RXJ1604.3-2130A}, henceforth J1604-2130) was first identified as a member of the Upper Scorpius OB association in the spectroscopic survey of X-ray selected sources of \cite{Preibisch:1999}.  We adopt the stellar parameters found by \cite{Preibisch:2002}: spectral type K2, stellar mass of 1.0 \Msun, extinction of A$_V =$ 1 mag, $\log$ T$_{eff}$ ~of 3.658, and $\log L/\Lsun$ ~of $-0.118$.  

 J1604-2130 is included in the UCAC3 proper motion survey \citep[][]{2010AJ....139.2184Z}.  They report a J2000 position of $\alpha_{2000}$ =16:04:21.6541 and $\delta_{2000}$ = $-$21:30:28.496, with errors on the J2000 position of 21 mas and 17 mas in right ascension and declination, respectively.  They also report a proper motion of -14.5$\pm$1.9 mas/yr in right ascension, -17.0$\pm$1.5 mas/yr in declination.

 Another association member, \object{RXJ1604.3-2130B}, lies 16" to the southwest (projected separation 2350 AU) and is a potential companion \citep{Kraus:2009}.  Aperture masking interferometry by \cite{2008ApJ...679..762K} ruled out the presence of a close binary companion of $\Delta K = 3.5$ mag at a distance of 10--20 mas, and $\Delta K = 6.2$ mag at 40--80 mas (corresponding to $\sim$ 0.06 \Msun ~at a distance of $\sim$2 AU, and $\sim$0.01 \Msun ~at $\sim$9 AU), while direct imaging in that work ruled out a binary companion as faint as $\Delta K \sim$5.5 at 240--320 mas (corresponding to $\sim$ 0.02 \Msun ~at a distance of $\sim$40 AU).  
The complementary study of \cite{Ireland:2011} has placed upper limits on companion masses of 0.07 \Msun ~to 0.005 \Msun at separations from 60 to $\geq$300 AU.

 High resolution optical spectroscopy shows evidence for very low levels of accretion, with the system exhibiting a P-Cygni profile in H$\alpha$ emission \citep[10\% full width of 252.7 km/s, equivalent width of  -0.57 -- -0.27 \AA, and heliocentric radial velocity of -6.27 -- -6.90 km/s; ][]{2009AJ....137.4024D,Dahm:2011}.  Recent Spitzer IRAC, IRS, and MIPS observations \citep{2006ApJ...651L..49C, 2009AJ....137.4024D, 2009ApJ...705.1646C} have revealed a transition disk SED, with no excess at 4.5 and 8 $\mu$m, and a rising excess at 16 $\mu$m and longer wavelengths.  \citealt{2009AJ....137.4024D} noted this distinctive SED, as well as evidence of a factor of 4 variability in the near-infrared excess indicating possible changes in the dust environment in the inner disk on short timescales.  \cite{Dahm:2010} presented J, H, and K band spectroscopy of this target, with weak H and K band excess consistent with blackbody emission at 900 K, suggesting the presence of dust at small orbital radii ($\leq$0.1 AU).    
 ~The 1.2 mm photometric survey of \cite{Mathews:2011} shows that this disk is the most massive in Upper Scorpius.

In \S \ref{sec:obs}, we describe our (sub)mm observations and data reduction.  We then show the continuum and CO line maps of the J1604-2130 transition disk, as well as analysis of the visibility data sets and mass estimates, in \S \ref{sec:results}.  Using a radiative transfer code, we model the dust component of the disk and estimate its parameters in Section \ref{sec:model}.  In \S \ref{sec:discussion} we discuss the implications for disc evolution and planet formation, and show that the disk has likely formed one or more giant planets.  We conclude with a summary in Section \ref{sec:summary}.

\begin{deluxetable*}{l l l l l l}
\tablewidth{0pt} 
\tablecolumns{6} 
\tablecaption{Observations}
\tablehead{
         \colhead{Date}	&  
         \colhead{Configuration}	& 
         \colhead{Bandpass}  &
         \colhead{t$_{int}$}	& 
         \colhead{$\tau_{225, zenith}$}	&	
         \colhead{median T$_{sys}$}
}
\startdata
            2008, July 11			& SMA Extended, 345 GHz		& 4 GHz 	&  2.0 hours	&	0.05	&  211 K   \\    
            2009, August 31		& SMA Extended, 345 GHz		& 8 GHz 	&  2.7 hours	&	0.09	&  331 K	\\ 
            2010, March 2		& SMA Very Extended, 345 GHz	& 8 GHz 	&  5.8 hours	&	0.03	&  139 K	\\ 
            2011, January 30		& PdBI A, 115 GHz				& 7.2 GHz	&  4.4 hours	&	0.17	&     466 K  	\\   
            2011, March 2		& PdBI B, 115 GHz				& 7.2 GHz	&  4.9 hours	&	0.12	&     418 K 	\\[-2mm]
\enddata
\label{tab:obs}
\end{deluxetable*}

\begin{figure*}[!ht]
\figurenum{1}
\centering
\includegraphics[width=6.5in]{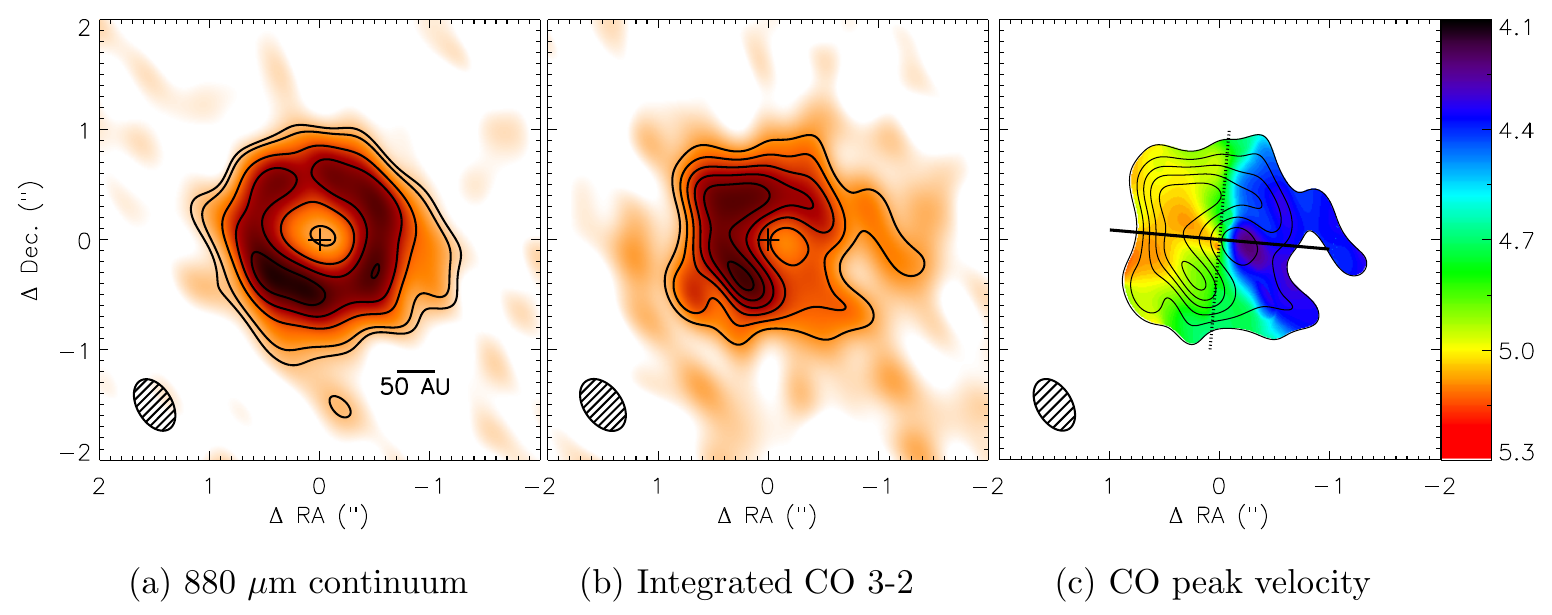}
\caption{\textbf{A)} 880 $\mu$m continuum map, with contours indicating 2, 3, 6, 9, and 12 $\sigma$ intensity (1$\sigma$ = 1.3 mJy/beam). 
\textbf{B)} Integrated CO J=3$\rightarrow$2 map from 4 to 5.5 km/s.  Contours indicate 2, 3, 4, 5, 6, and 7$\sigma$ emission, in 1$\sigma$ steps (1$\sigma$ = 75 mJy km/s /beam ).
\textbf{C)}  Peak CO J=3$\rightarrow$2 velocity map, limited to regions with 2$\sigma$ integrated line emission. 
In a) and b), the stellar position at 16:04:21.645 -21:30:28.66 is indicated with a cross, and the solid and dashed lines in c) indicate the directions of the major and minor axes, respectively.  The $\sim0\farcs5 \times 0\farcs3$ beam is shown in the bottom left. }
\label{fig:CO-map}
\end{figure*}

\bigskip
\bigskip
\section{Observations}
\label{sec:obs}

 Interferometric observations of J1604-2130 were carried out in three tracks with the Sub-Millimeter Array \citep[SMA, ][]{Ho:2004} and in two tracks with the Plateau de Bure interferometer \citep[PdBI, ][]{Guilloteau:1992}.  In Table \ref{tab:obs}, we give the night, array configuration, total continuum bandwidth, total time on-source, zenith sky opacity at 225 GHz, and median system temperature for each observation. 

  SMA observations consist of upper and lower bandpasses of 4 GHz bandwidth (2 GHz prior to 2009) for a total continuum bandwidth of 8 GHz (4 GHz prior to 2009), with the centers of each bandpass separated by 5 GHz and a local oscillator frequency of 340.4 GHz (880 $\mu$m).  Continuum regions were sampled at 3.25 MHz, and a 104 MHz wide region was set aside at the rest frequency of the CO J=3$\rightarrow$2 line (345.79599 GHz) to carry out higher spectral resolution observations with a sampling of 0.203 MHz.   Extended array observations included baselines from 44 to 226 m (50 -- 257 k$\lambda$), and very extended array observations included baselines from 120 to 509 m (136 -- 578 k$\lambda$).  Doppler tracking was carried out in the kinematic LSR rest frame.  Our observations were carried out in generally good to excellent conditions, with system temperatures of $\sim$100 -- 400 K, and zenith opacities at 225 GHz ($\tau_{225}$) less than 0.1.

 For gain calibration, we interleaved alternate 4--5 minute observations of the quasars QSO B1514-24 and QSO B1730-130 between  20--26 minute observations of J1604-2130.  For bandpass calibration, we used the bright quasars 3C454.3 or 3C273, and Uranus, Callisto, or Titan for flux amplitude calibration (as available).  Observations of flux and bandpass calibrators were carried out before or after J1604-2130 was available.  

 Using standard routines in the facility IDL package MIR, we flagged and calibrated the data.  We carried out baseline based phase calibration, finding rms phase errors of $\sim$10--20\degr.  Based on variations in the measured fluxes of our gain calibrators, the flux calibration has an uncertainty of $\sim$15\%.  

 We combined wideband continuum channels from all observations and both sidebands in the imaging process.  We used the facility reduction tool MIRIAD to carry out Fourier inversion, CLEAN deconvolution and image reconstruction using natural weighting.    
The synthesized beam of the combined observations is $0\farcs51 \times 0\farcs32$, with an rms of 1.3 mJy/beam in the continuum.  We separately carried out the same imaging process on the higher spectral resolution line data, binning in 0.25 km/s velocity channels centered on the CO J=3$\rightarrow$2 line.  The beamsize is $0\farcs53 \times 0\farcs34$, and the median noise in the channels is 125 mJy/beam.  

 PdBI observations were carried out on two nights, January 30, 2011 and March 2, 2011, in the A and B configurations, respectively.  Continuum observations at 2.6 mm used the WideX receiver with 3.6 GHz upper and lower sidebands for a total bandpass of 7.2 GHz and sample spacing of 1.95 MHz.  Line observations were carried out in a 20 MHz wide region at the rest frequency of the CO J=1$\rightarrow$0 line (115.271 GHz), with a sampling of 0.039 MHz.  The A configuration observation included baselines from 136 to 760 m (52 -- 292 k$\lambda$), while the B configuration observation included baselines from 88 to 452 m (34 -- 174 k$\lambda$).  System temperatures ranged from 400 to 800 K, and zenith $\tau_{225}$ ranged from  0.12 to 0.17 (corresponding to median water vapor column from 2.3 to 3.5 mm PWV).

 Gain calibration was carried out using 3--6 minute observations of 1504-166, 1519-273, and 1657-261 (as available) between 25 minute observations of J1604-2130, while flux and bandpass calibration were carried out with observations of MWC349 and 3C279.  We used the facility reduction tool GILDAS to carry out data calibration, Fourier inversion, and imaging.  The combined continuum observations have a beam size of $2\farcs01 \times 0\farcs70$, with an rms of 0.25 mJy/beam in continuum.    
From repeated observations of the calibrators, we estimate a flux calibration uncertainty of 20\%.

\begin{figure*}[!ht]
\figurenum{2}
\centering
\includegraphics[width=6.5in]{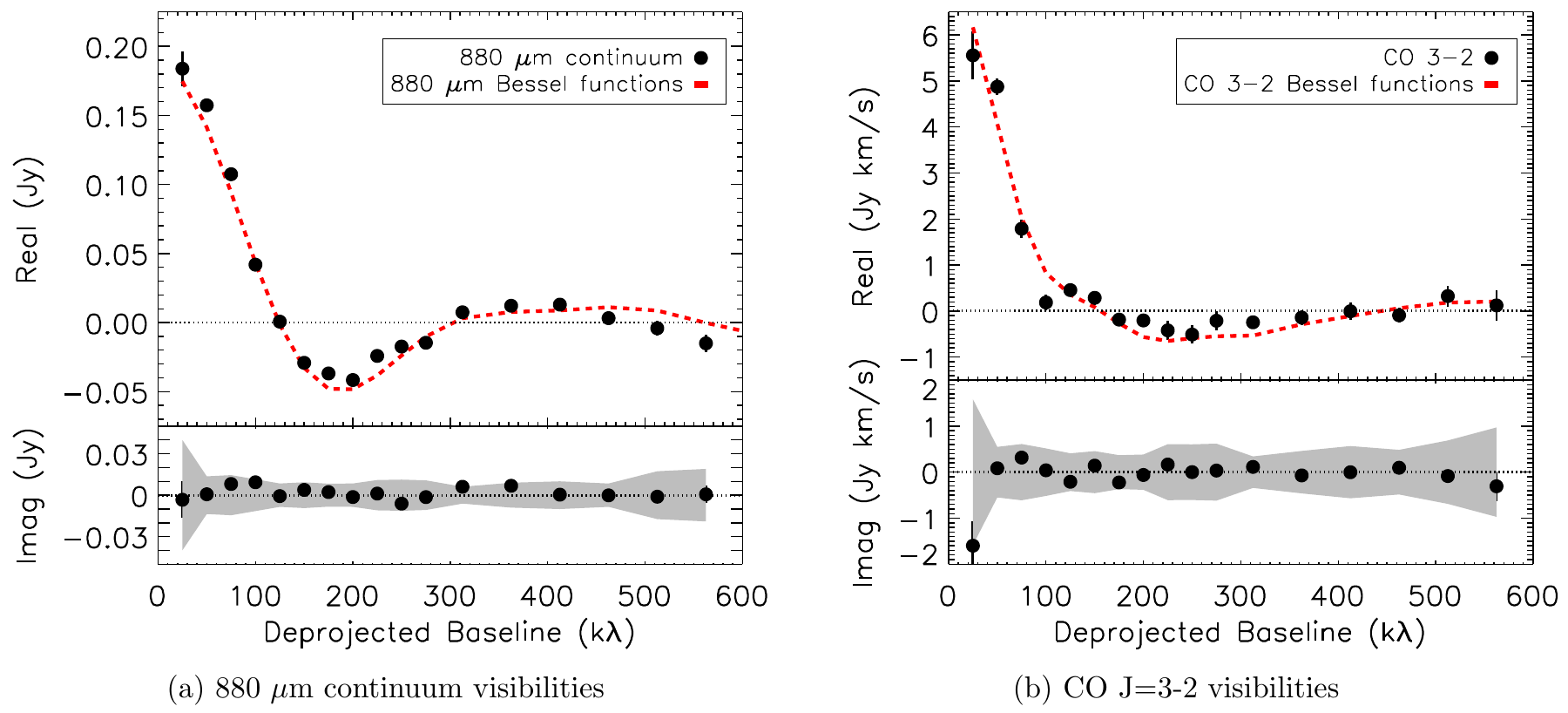}
\caption{Real flux components of the continuum and CO J=3$\rightarrow$2 visibilities.  1$\sigma$ errors are shown as vertical bars, and model fits \citep[i.e. integrated zeroth order Bessel functions of the first kind, following][]{2007ApJ...664..536H} shown as dashed red lines.  The range of 3$\sigma$ values for the imaginary components are shown as the gray region.  The position of the first null strongly constrains the radius of the inner edge of the circumstellar disk (50 -- 90 AU, for varying radial emission functions).}
\label{fig:uvfigs}
\end{figure*}

\section{Results}
\label{sec:results}

We discuss first the appearance of the SMA maps, and then analyze the velocity map and continuum visibilities, along with continuum photometry from the literature, to determine the disk properties.  In the PdBI maps, the disk is barely resolved in the east-west direction.  As the SMA data is at much higher resolution, we use the PdBI continuum measurement as a point in the SED.  From the continuum data, we measure the total flux at 880 $\mu$m and 2.6 mm.  From the integrated emission maps we measure the total flux in the CO J=3$\rightarrow$2 and J=1$\rightarrow$0 lines, and from the CO J=3$\rightarrow$2 peak-velocity map we determine the orientation of the disk.  We then use the continuum radially averaged visibilities to infer the central position and the inner radius of the disk.  These parameters are then refined via modeling, in Section \ref{sec:model}.  We list the properties derived in this section in Table \ref{tab:prop}.

\begin{deluxetable}{l l}
\tabletypesize{\scriptsize}
\tablewidth{0pt} 
\tablecolumns{2} 
\tablecaption{Observational Properties}
\tablehead{
\colhead{Property}	&  
\colhead{Value}	\\ 
}
\startdata
	F$_{880\mu m}$ 			&	164$\pm$6 mJy \\
	F$_{CO~J=3\rightarrow2}$ 	&	$5.2\pm0.1$ Jy km/s \\
	F$_{2600 \mu m}$ 				&	5.1$\pm$0.5 mJy \\
	F$_{CO~J=1\rightarrow0}$ 	&	$0.48\pm0.04$ Jy km/s \\
	central position					&	16:04:21.645 -21:30:28.83  \\
	inclination						&	6\degr$\pm$1.5\degr \\
	position angle					&	-5\degr$\pm$10\degr  \\
	systemic velocity (LSRK)			&	4.7 km/s $\pm$0.1 km/s   \\[-2mm]
\enddata
\label{tab:prop}
\end{deluxetable}

\subsection{SMA images}
\label{sec:SMAimages}

We show the 880 $\mu$m continuum emission map for J1604-2130 in Figure~\ref{fig:CO-map}, centered on the disk central position.  The disk is being viewed from a nearly face-on perspective, and has a large inner hole.  The image has a peak flux of 18.2 mJy/beam.  To measure the total emission, we integrate all flux above the 2$\sigma$ contour, giving  a total flux of 165$\pm$6 mJy.

\label{sec:COmap}

CO J=3$\rightarrow$2 line emission was found at 2$\sigma$ and greater significance in channels from 4.0 and 5.5 km/s (LSRK).  We sum these 0.25 km/s wide channels to produce the integrated intensity map seen in Figure \ref{fig:CO-map}, with an rms of 75 mJy km/s / beam.  The integrated intensity map shows a drop in emission to the southwest and in the center of the disk; however, the emission does not clearly drop below 3$\sigma$ in either location.  The outer radius of the CO emission roughly corresponds to that of the continuum emission.  Integrating emission above the 2$\sigma$ contours of integrated line emission, we find a total CO J=3$\rightarrow$2 line flux of $5.2\pm0.1$ Jy km/s.  Following the same procedure, we find a line flux of $0.48\pm0.04$ Jy km/s for the CO J=1$\rightarrow$0 data.

We generated the peak velocity map by fitting a gaussian emission profile at each spatial position in the disk, including emission from 0.75 to 8.5 km/s in order to achieve good baseline fits and ensure we captured the full profile at each location.  The median uncertainty in the peak velocity at each point is 0.07 km/s.  This map shows a clear east-west velocity gradient which we use to measure the position angle and inclination of the disk.  We do this by fitting a Keplerian velocity profile (Figure~\ref{fig:CO-map}).  Velocities along the observed minor axis of the disk should be constant, reflecting the systemic velocity.  Velocities along the major axis will be proportional to $r^{-1/2}$sin$i$, where $r$ is the distance from the star and $i$ is the inclination from our line of sight (with $i=0$ indicating a face-on disk).  

To deduce the observational geometry, we varied the central position, position angle, and inclination, calculated line-of-sight velocities along the major and minor axes, and carried out a $\chi^2$ fit to the velocity map.  We mask the central 0\farcs53 (the beam major axis) of the velocity map, since beam dilution will mask the large range in velocities expected at small radii and lead to poor fits.  We find the disk is centered at $\alpha_{2000}$ = 16h04m21.6416s and $\delta_{2000}$ = $-$21\degr30'28\farcs798, with a conservative uncertainty of $\sim0\farcs1$ in both right ascension and declination.  The predicted position from the UCAC3 position and proper motion is 16h04m21.6394s, $-$21\degr30'28\farcs661.  Taking into account the J2000 position uncertainty, proper motion uncertainty, and the position uncertainty of our observations, the offset between the predicted stellar position and the disk center is $\Delta\alpha = 0\farcs031\pm0\farcs104$ and $\Delta\delta = 0\farcs137\pm0\farcs102$, respectively.  The disk is inclined at $6\pm1.5$\degr ~from face on, at a position angle of 5\degr$\pm$10\degr ~west of north, and with a central velocity of 4.7 $\pm$ 0.1 km/s.  

We use these parameters for modeling the dust disk, but defer further quantitative discussion of the line data to a later paper.

\subsection{Radially averaged visibilities}
\label{sec:uvprofile}

Based on the disk inclination and position angle as determined from the peak velocity map, we deproject the interferometer visibilities, scaling uv distances along the minor axis by the cosine of the disk inclination.  We plot the radially averaged real and imaginary fluxes versus the radial uv distance.  To provide an adequate signal-to-noise ratio, we bin the real and 
imaginary flux in 25 k$\lambda$ bins from 12.5--287.5k$\lambda$, and 50k$\lambda$ bins at larger baselines.

We plot the binned components of the continuum visibilities in Figure \ref{fig:uvfigs}, along with the similarly binned CO J=3$\rightarrow$2 visibilities, including measurements from 4 to 5.5 km/s as in the integrated emission map.  The imaginary component is less than 3$\sigma$ in all bins, making it consistent with zero as expected for a centered, azimuthally symmetric disk.   We find that a greater fraction of the power in the CO line is found at small baselines compared with the continuum emission, suggesting more CO emission is to be found in large scale structures.

The radially averaged profiles strongly constrain the size of the inner hole in gas and dust, and provides further insights to the large scale structure of the disk.  Strong contrast between the cavity and outer disk leads to nulls in the Fourier transform of the emission profile.  To estimate the inner and outer edges of sub-millimeter emission, we fit a simple disk model to the radially binned real visibilities based on Equation A2 of \cite{2007ApJ...664..536H}:

\begin{equation}
  Vis(R_{uv}) = A \int_{\theta_1}^{\theta_2} \theta^{1-p} J_0 (2\pi\theta R_{uv}) \,\mathrm{d}\theta., 
\end{equation}
which represents the Fourier transform of a power-law brightness disk with an inner hole (i.e. $I(\theta) \propto \theta^{-p}$).  $A$ is a free parameter that sets the intensity scale, $\theta_1$ and $\theta_2$ are the inner and outer disk edges respectively, and $p$ is the exponent of a power law describing the variation in emission intensity with radius.  $J_0$ is the Bessel function of the first kind of order 0.  We show our best-fit analytic models in Figure \ref{fig:uvfigs}.  The continuum model has parameter values of $A = 0.16\pm0.02$ Jy, $\theta_1 = 0\farcs43\pm0\farcs01, \theta_2 = 0\farcs95\pm0\farcs03,$ and $p = 2.03\pm0.01$.  The CO line model has parameter values of $A = 5.0\pm0.2$ Jy km/s, $\theta_1 = 0\farcs32\pm0\farcs01, \theta_2 = 1\farcs84\pm0\farcs24,$ and $p = 1.98\pm0.01$.  The similar values for $p$ ($\sim 2$) suggest the brightness functions of dust and gas have similar variation with radius.

The inner and outer angular limits correspond to radii from 60 -- 140 AU for continuum emission, and 45 -- 270 AU for CO emission.  These suggest the dust cavity may be larger than that of the gas, however, given the complex interaction of surface density and temperature structure in determining the surface brightness function, such an assessment cannot be made without more detailed modeling.  We also note that the continuum model does not fit the long baseline visibilities well, suggesting the presence of some millimeter continuum emission at small spatial scales.  We examine the structure of the dust disk with detailed radiative transfer modeling, including the spectral energy distribution, but we defer further interpretation of the CO line data to a later paper.

\subsection{Millimeter-emitting dust mass}
\label{sec:mass}

Under the assumption of optically thin dust emission, we can estimate the dust mass of the disk.  The dust mass ($M_{dust}$) will be directly proportional to the sub-mm flux (F$_{\nu}$), as in \cite{Hildebrand:1983}:

\begin{equation}
M_{dust} = \frac{d^2F_{\nu}}{\kappa_{\nu}B_{\nu}(T_c)}
\end{equation}

We assume J1604-2130 lies at the mean distance of Upper Sco derived from the Hipparcos data \citep[145 pc,][]{de-Zeeuw:1999}.  We also use a characteristic dust temperature ($T_c$) of 20 K as shown to be appropriate for Taurus disks \citep{2005ApJ...631.1134A}, and a dust mass opacity $\kappa_{\nu}$ = 10 ($\nu$/1000 GHz)$^{\beta}$ cm$^2$/g \citep{Beckwith:1990}.  

\begin{figure}[tb]
\figurenum{3}
\centering
\includegraphics[height=2.5in]{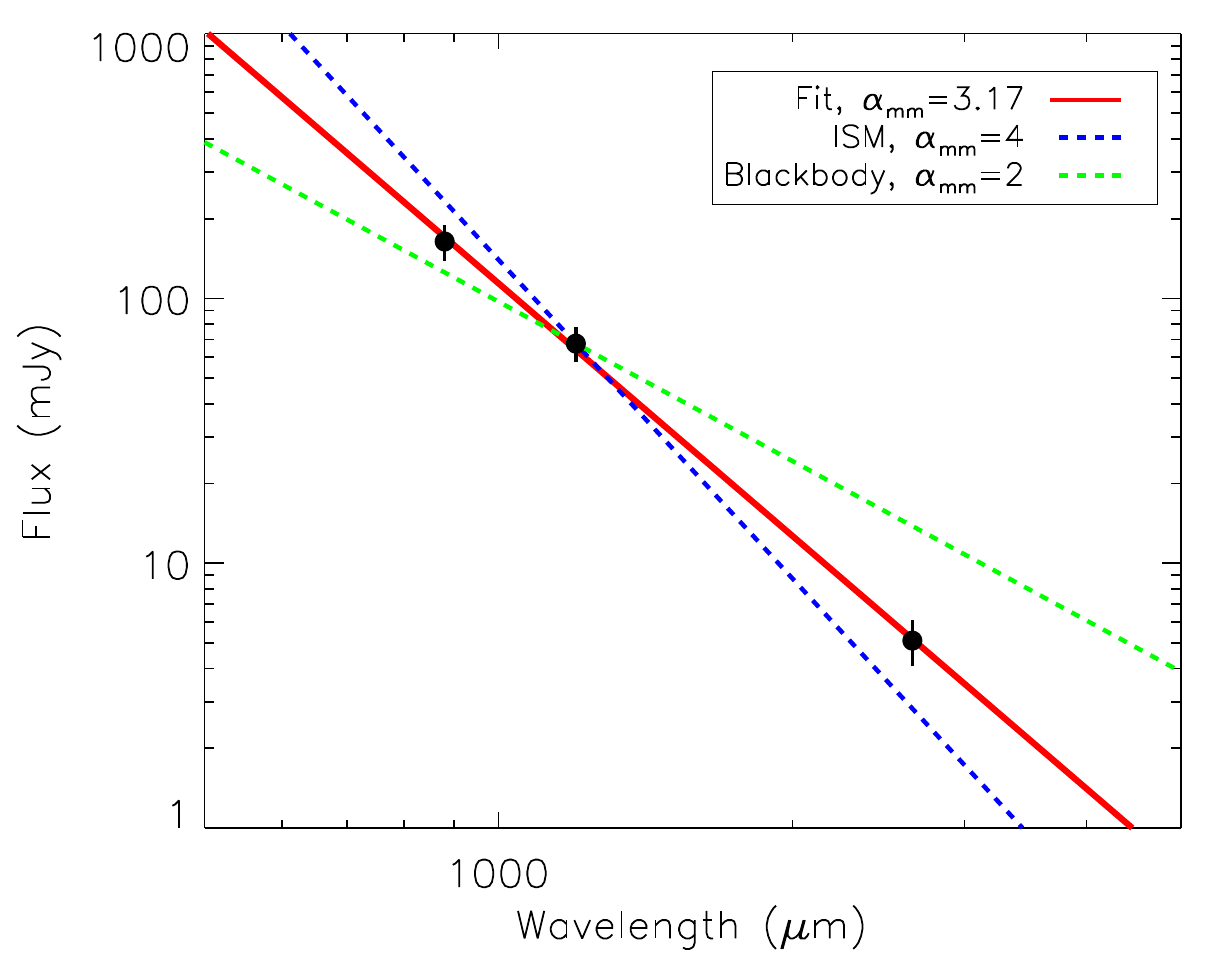}
\caption{We show the millimeter SED as black dots, along with power law fit and comparisons to $\alpha_{mm}$ = 2 and 4, corresponding to opacity power laws of $\beta = 0$ and $\beta = 2$ in the case of optically thin emission.  Errors include both formal and calibration uncertainty.}
\label{fig:mm-sed}
\end{figure}

We estimate the value of $\beta$ by making a power law fit to the mm-fluxes.  We plot the 880 $\mu$m and 2.6 mm fluxes in Figure \ref{fig:mm-sed}, along with the 1.2 mm flux from \cite{Mathews:2011}, $67.5\pm1.4$ mJy.  A power law fit, including calibration uncertainties, shows that for $f_{\nu} \propto \nu^{-\alpha_{mm}}$, the mm-slope $\alpha_{mm}$ is equal to $3.17\pm0.21$.  Assuming an optically thin medium and emission in the Rayleigh-Jeans limit, where $\alpha_{mm} = \beta + 2$, we then calculate a value of $\beta = 1.17$.  This is approximately 3$\sigma$ less than the value of $\beta$ for small ISM grains ($\beta = 2$), and is greater than that expected for large blackbody grains ($\beta = 0$).  It is close to the value of $\beta \approx 1$ expected in the case where grains follow a power law size distribution $N(a) \propto a^{-3.5}$, with a maximum grain size at least 3 times the observed wavelength \citep{Draine:2006}.  Together, these suggest that this disk may have experienced significant grain growth, up to at least centimeter-sizes.  The inferred dust mass is 0.1 $\,M_{\rm Jup}$.  

The assumption of optically thin emission is justified because, for the $\beta = 1.17$ opacity function, dust will become optically thick to 880 $\mu$m emission at a surface density of 0.35 g/cm$^2$, and at a surface density of 1.3 g/cm$^2$ for 2.6 mm emission.  Surface densities this high are not observed at large radii in even the youngest primordial disks.  For our estimated disk mass, this high density would require a 2--5 AU wide ring at a radius of 90 AU.  However, such a narrow ring is ruled out by both the SMA images and the simple model fits described above.  In addition, we show in our later modeling (\S \ref{sec:fit}) that the peak surface density in the disk is a factor of ten too low for the dust to be optically thick to 880 $\mu$m emission.  

\section{Modeling}
\label{sec:model}

The temperature profile of the disk is reflected in the spectral energy distribution (SED), while the density profile is reflected in the resolved sub-mm image.  By simultaneously fitting the SED and SMA data, we derive the disk properties.  Because of the sparse sampling of the Fourier plane, we fit the interferometric visibilities rather than the image itself.  For the stellar properties, we adopt a stellar temperature of 4550 K, luminosity of 0.76 \Lsun, $\log g$ of 4.0, and extinction of 1 mag in V band, based on the references in \S\ref{sec:intro}.  For the disk orientation, we use the disk position angle and inclination derived in Section \ref{sec:SMAimages}.

\subsection{Procedure}
\label{sec:procedure}

To model the disk, we use the 3-dimensional radiative transfer code MCFOST \citep{Pinte:2006,Pinte:2009} to produce a model SED and 880 $\mu$m image.  We then simulate observation of that image to generate visibilities.  We fit the SED and radially binned components of the 880 $\mu$m continuum visibilities.

We constructed a spectral energy distribution for J1604-2130, combining our 880 $\mu$m flux with values in the literature.  For the B, R, and I bands, we used photometry from \cite{Preibisch:1999}, converting magnitudes to fluxes using the absolute flux conversions of \cite{Bessell:1979}.  We used near-infrared J, H, and K$_S$ band photometry from the 2MASS survey \citep{Skrutskie:2006} and the absolute flux conversions of \cite{Cohen:2003}.  The near-infrared 4.5, 8, and 16 $\mu$m photometry of \cite{2006ApJ...651L..49C} and mid-infrared 24 and 70 $\mu$m photometry of \cite{2009ApJ...705.1646C} were published as fluxes; these values are used here directly.  We also include the spatially integrated 880 $\mu$m flux from this work, and the single-dish 1.2 mm continuum measurement from \cite{Mathews:2011}.  We dereddened the optical and near-infrared fluxes by A$_V$ = 1 mag using the R=3.1 reddening law of \cite{Fitzpatrick:1999}.  

The SMA image indicates a disk with a large inner hole.  In addition, the SED shows only photospheric emission out to 8 $\mu$m, requiring an absence of dust out to several AU.  However, the excess emission at 16 and 24 $\mu$m requires the presence of a small amount of dust within the inner hole seen in the SMA image.  A model that has an inner hole entirely depleted of dust therefore does not fit the data, and we require a three zone disk model.

Following \cite{Andrews:2011c}, we explore a modified version of the parametric accretion disk density model with a dust surface density of

\begin{equation}
  \Sigma_{acc} (R) = \Sigma_C \left ( \frac{R}{R_C} \right ) ^{-\gamma} \exp \left [-\left (\frac{R}{R_C} \right )^{2-\gamma} \right ], 
\end{equation}
where $\Sigma_C$ is the surface density at the characteristic scaling radius, $R_C$, and $\gamma$ is treated as a free parameter describing the surface density of the disk \citep[this corresponds to the viscosity power law index in accretion disk theory, e.g.][]{Hartmann:1998}.  

We split the disk into three regions.  In the innermost part of the disk, within a gap of radius $R_{gap}$, we treat the dust surface density as zero to account for the lack of excess at wavelengths up to 8 $\mu$m.  Based on the hole seen in 880 $\mu$m emission, yet accounting for the excess seen at 16 and 24 $\mu$m, we deplete the inner region of the disk by a factor $\delta_{cav}$ from $R_{gap}$ to $R_{cav}$.  We refer to this depleted region as the disk cavity, and to the remainder as the disk.  For the outermost radius included in our models, we set $R_{out}$ to 300 AU, but in practice, the emission at these radii is too low for the sensitivity of our data.  Therefore, our adopted surface density model goes as:
\begin{equation}
  \Sigma_{dust} (R) =  \left\{
  		\begin{array}{r l} 
                               	0, 																							& 	\mbox{if } R < R_{gap} \\
                               	\delta_{cav}  \Sigma_{acc} (R),	&	\mbox{if } R_{gap} \le R < R_{cav} \\
                               	 \Sigma_{acc} (R),				&	\mbox{if } R_{cav} \le R < R_{out} 
		\end{array}
		\right.
\end{equation}

The vertical dust distribution is modeled as a Gaussian with scale height $h=h_C R_C$ $(R/R_C)^{\psi}$, where $h_C R_C$ is the scale height at $R_C$ and $\psi$ describes the power-law disk flaring.  After initial modeling indicated our data do not constrain $\psi$, we adopted a value of 1.2 for this parameter, based on the median flaring value found in the multi-transition disk study of \cite{Andrews:2011c}.  

For the dust properties, we adopt the porous grain mineralogy of \cite{Mathis:1989}, fixing the exponent of the power law grain size distribution at $-3.5$.  We adopt a minimum grain size of 0.005 $\mu$m and a maximum grain size of 7.8 mm \citep[i.e. 3 times the longest observed wavelength, following][]{Draine:2006}.  The grain composition and grain size distribution determine the wavelength-dependent opacity function.  We also note that the choice to use the grain properties of \citeauthor{Mathis:1989} is essentially arbitrary.  Other grain models lead to equally good models and the current data on J1604 does not discriminate among the several possibilities.  

In order to find a best-fit model for the J1604-2130 disk, we carried out a search of the 7 free parameters identified above ($\Sigma_C$, $R_C$, $\gamma$, $R_{gap}$, $R_{cav}$, $\delta_{cav}$, $h_C$), adopting the central position, inclination, and position angle determined from the peak-velocity map (\S \ref{sec:COmap}).  Generating even a sparse grid of 7 parameters would require a prohibitively long time, therefore, we use the Levenberg-Marquardt $\chi^2$ minimization algorithm as implemented in the IDL routine MPFIT \citep{Markwardt:2009} to carry out an iterative least squares fit to the observed SED and real and imaginary components of the radially averaged visibilities.  Each evaluation of the parameter search carries out the following steps:

\begin{enumerate}
\item For a given set of parameters, we generate the corresponding model at the observed central position, position angle, and inclination.  MCFOST produces both an SED for the model and an 880 $\mu$m image.

\item We then simulate an observation of the model with the uv-coverage and integration time of our SMA observations.  We use the Fourier transform functionality of the CASA package\footnote{http://casa.nrao.edu/} to generate a simulated uv dataset for the model image, from which we generate deprojected, recentered, and binned radially averaged visibilities.    

\item The fitting routine calculates the local numerical gradient of the $\chi^2$ function of the SED and the radially averaged visibilities.   As the model disk is azimuthally symmetric, the imaginary components of the radially averaged model visibilities are always zero, with slight variations due to the stochastic nature of the image simulation.

\item The numerical gradient of the $\chi^2$ values are then used in determining the next point to sample in the space of free parameters.  The search algorithm follows the $\chi^2$ gradient towards the minimum.  

\end{enumerate}

In order to avoid local minima and more fully map the parameter space, we carried out the search algorithm 40 times, randomly varying the starting values of the parameters within the ranges listed in Table \ref{tab:models}.  The Levenberg-Marquardt algorithm naturally leads to low density sampling in regions of the parameter space with high $\chi^2$ values, with the sampling density rising as the search moves into regions of lower $\chi^2$.  In addition to providing a sampling of the parameter space, each run, $k$, of the search algorithm results in a set of best-fit values for the $i$ parameters ($\mu_{k}  = \mu_{1,k}, \mu_{2,k}, ..., \mu_{i,k}$), parameter uncertainty estimates ($\sigma_{k} = \sigma_{1,k}, \sigma_{2,k}, ..., \sigma_{i,k}$), and a covariance matrix ($A_k$) which allows for estimation of the parameter dependencies.

\newpage
\subsection{Fit Results}
\label{sec:fit}

\begin{deluxetable}{l c c r l}
\tabletypesize{\scriptsize}
\tablewidth{0pt} 
\tablecolumns{5} 
\tablecaption{Dust parameters for USco J1604}
\tablehead{
         \colhead{Parameter}	&  
         \colhead{Mean}	& 
         \colhead{Standard} 	& 
         \colhead{Search} &
         \colhead{}  	\\ 
         \colhead{ }		&
         \colhead{value}		         &
         \colhead{deviation} &
         \colhead{Lower}         &
         \colhead{Upper}
}
\startdata
			\hline %
			log($\Sigma_c$) [$g/cm^2$]	&  -1.23	&  0.17	&  -3		&  1			\\  	
			$R_C$ [$AU$]				&  106	&  7		&  90		&  150		\\  	
			$\gamma$				&  -0.40	&  0.43	&  -2.0	&  2.0		\\  	
			$R_{gap}$ [$AU$]			&  20		&  3		&   15 	&  49			\\    	
			$R_{cav}$ [$AU$]			&  72		&  3		&    50	&  100		\\    	
			log($\delta_{cav}$)			&  -0.86	&  0.35	&   -3 	&  0			\\    	
			$h_C$					&  0.04	&  0.01	&    0.010	&  0.100		\\    	
			$M_{dust} (M_{Jup})$		& 0.13	&  ---		&  --		&  --			\\[-2mm]
\enddata
\label{tab:models}
\end{deluxetable}

We carried out 40 parameter searches starting from very different initial values, each of which consisted of about 100 disk models. The model with the lowest $\chi^2$ value out of the total of 3967 separate disk models has $\chi^2 = 75.7$ and parameters of log $\delta_{Cav} = -0.89$, $R_C = 108 $AU, $\gamma = -0.69$, log$\Sigma_C = -1.23$ g/cm$^2$, $h_C = 0.04$, $R_{gap} = 19$ AU, and $R_{cav} = 70$ AU.  35 of the 40 searches converged to a very similar set of parameters, with $\chi^2$ ranging from 76 to 85.  This gives us confidence that we are reaching the global $\chi^2$ minimum of the seven dimensional parameter space.  The remaining 5 of the 40 searches converged on local minima with $\chi^2$ greater than 100.

The separate searches show strong convergence, despite starting in different regions of the parameter space.  These models tend to have similar $\chi^2$ values, suggesting that a mean of the parameters found by the several searches will provide a reasonable estimate of the disk parameters.  We make use of the full set of $\sim4000$ disk models to map out parameter space so we can assess the reliability of our parameter estimates.  

There are, however, difficulties in estimation of the parameter uncertainties.  Formally, our models have 40 degrees of freedom (i.e. 14 SED points, 17 visibility bins with both real and imaginary components, and 7 free parameters), for a minimum reduced $\chi^2$ value of $\approx2$.  In addition, some data points do not contribute to the fitting process (e.g. optical photometry, where the disk cannot have emission, and imaginary visibility points which will be near zero for all azimuthally symmetric models), and others will be highly correlated (e.g. photometry near the SED peak and visibilities near the first null, both of which are strongly affected by the inner disk radius) and thus will not contribute a full degree of freedom.  The combination of these two issues -- possible model misspecification and uncertainty in the number of degrees of freedom --  suggests that the uncertainty estimates produced by MPFIT for each search and the conversion of $\chi^2$ values to probabilities are inaccurate.

This means that traditional approaches to error estimation (e.g. $\Delta\chi^2$ contours) are not applicable.  However, the relative comparison of models via differences in $\chi^2$ are still valid, i.e. a lower $\chi^2$ value still indicates a better model.  We cannot say, however, to what degree the model is better.  With these limitations in mind, we report the standard deviation of best-fit parameter values, but note that they do not necessarily represent 1$\sigma$ uncertainties, and only serve to give a qualitative indication of how well each parameter is constrained in relation to the range of values included in the search.  We report the mean parameter values and their standard deviations, along with the range of values explored for each parameter, in Table \ref{tab:models}.

\begin{figure}[b]
\figurenum{4}
\centering
\includegraphics[height=3in]{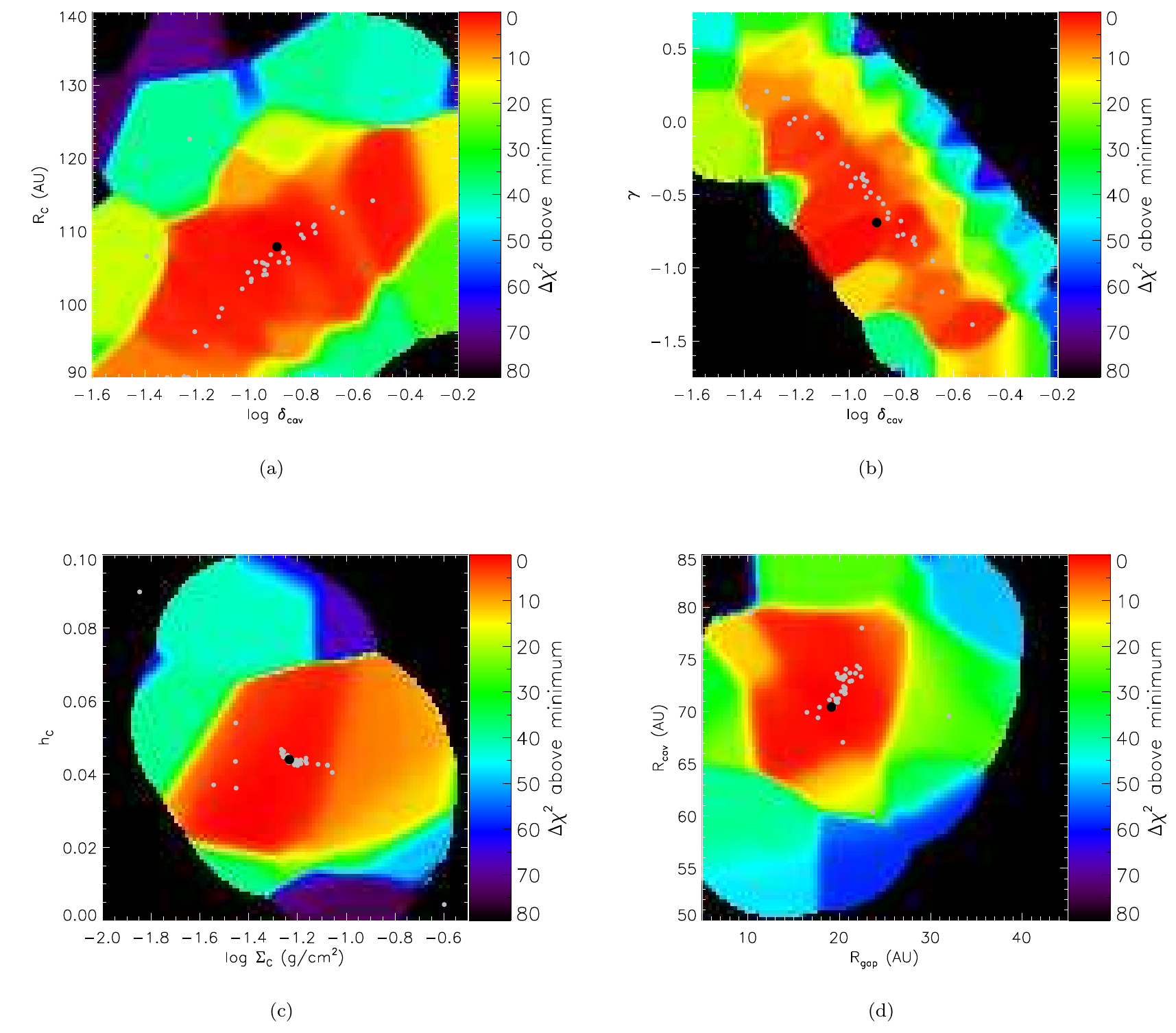}
\caption{2-dimensional projections of the 7-dimensional $\chi^2$ distribution ( $\delta_{Cav}$ appears twice, to highlight the strong covariance with $R_C$ and $\gamma$).  Gray circles show the results of the searches for the bestfit parameters and the black circle shows the parameters of the minimum-$\chi^2$ model.  The apparent blocky structures in the high $\Delta\chi^2$ regions are artifacts of sparse sampling and high $\chi^2$ gradients.  The projection of the low $\Delta\chi^2$ region to the left in Figure b is an artifact of the distance contribution term, as the minimum $\chi^2$ model contributes disproportionately to this nearby high $\chi^2$ region, which has a low sampling density.}
\label{fig:chisq}
\end{figure}

For a further qualitative examination, we show the 2-dimensional $\Delta\chi^2$ distributions for pairs of parameters (Fig. \ref{fig:chisq}).  At each point within the displayed space, we calculate the parameter-distance and $\chi^2$-weighted mean $\chi^2$.  In each figure, we show best-fit parameters from our searches.  
The inner disk dust depletion factor, $\delta_{cav}$, strongly correlates with both $R_C$ and $\gamma$, the characteristic radius and surface density power law term.  The other parameters ($\Sigma_C$, $h_C$, $R_{cav}$, and $R_{gap}$) are tightly constrained with only weak correlations with other parameters.

\begin{figure*}[!ht]
\figurenum{5}
\centering
\includegraphics[width=6.5in]{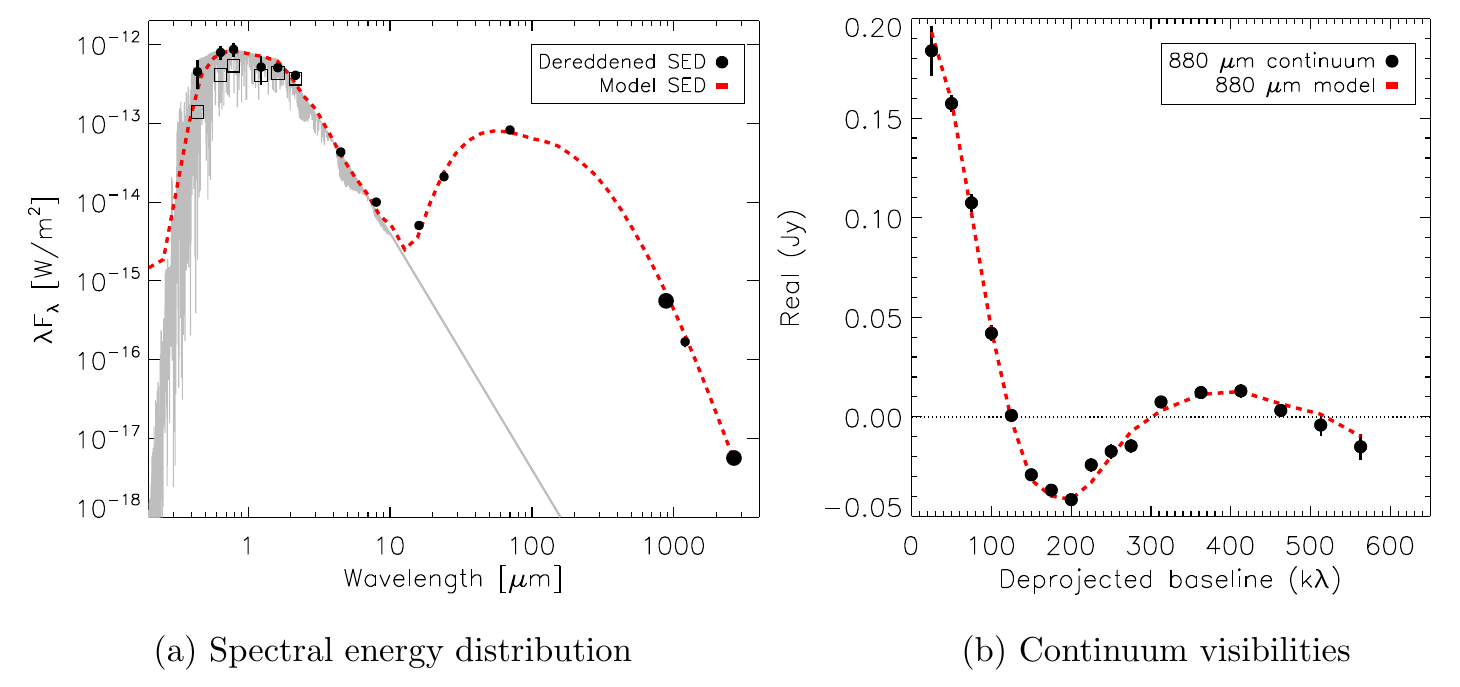}
\caption{Model fits to SED and continuum visibilities.
A)  Spectral energy distribution for J1604-2130.  We show our 880 $\mu$m and 2.6 mm fluxes as large black circles, and dereddened optical, 2MASS, Spitzer, and 1.2 mm IRAM 30m observations as solid black circles.  We assume calibration errors of 20\% for optical photometry, 10\% for infrared photometry, 15\% for 880$\mu$m and 1.2mm photometry, and 20\% for 2.6mm photometry.  Original photometry at short wavelengths is shown with black squares.  The closest match stellar model from \cite{Kurucz:1979,Kurucz:1993} is shown as a gray line.
B) Real flux components of the 880 $\mu$m continuum as a function of deprojected uv distance.  1$\sigma$ errors are shown as vertical bars. }
\label{fig:modfits}
\end{figure*}

\begin{figure*}[!ht]
\figurenum{6}
\centering
\includegraphics[width=6.5in]{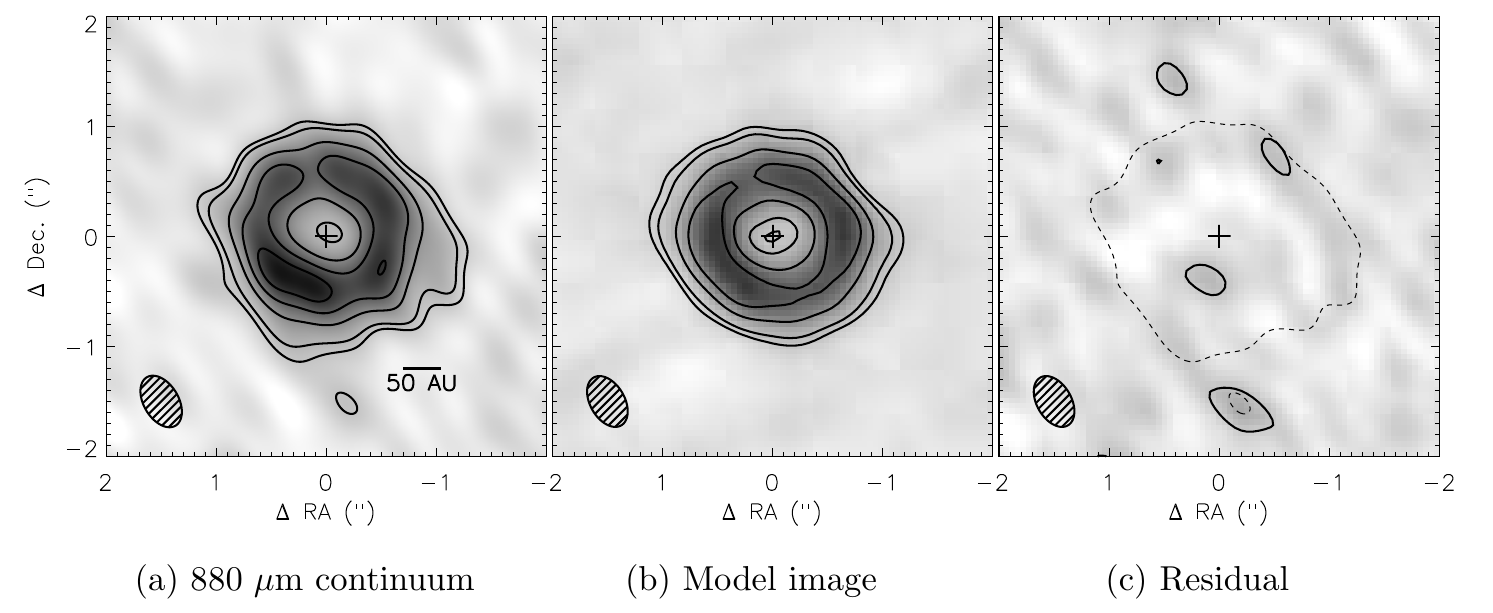}
\caption{\textbf{A)} 880 $\mu$m continuum map, with contours indicating 2, 3, 6, 9, and 12$\sigma$ intensity.
\textbf{B)} 880 $\mu$m model image, with contours indicating 2, 3, 6, 9, and 12$\sigma$ intensity.
\textbf{C)}  880 $\mu$m residual map, with solid contours indicating 2$\sigma$ residuals, and the dashed line indicating the 2$\sigma$ contour of the observed continuum map.
In all images, the stellar position at 16:04:21.645 -21:30:28.66 is indicated with a cross, and 1$\sigma$ = 1.3 mJy/beam.}
\label{fig:contmodel}
\end{figure*}

We show the comparison of our adopted disk model with the data in Figures \ref{fig:modfits} (SED and binned visibilities) and \ref{fig:contmodel} (continuum image, model, and residual).  We calculate the residual by subtracting the model visibilities from the observed visibilities, and then carrying out the same imaging procedure as described in \S{\ref{sec:obs}}.  We can see that the model underpredicts the flux at the 2$\sigma$ level at the location of the peak of emission to the southeast, suggesting a possible asymmetry in the continuum emission.  

\begin{figure}[b]
\figurenum{7}
\centering
\includegraphics[height=2in]{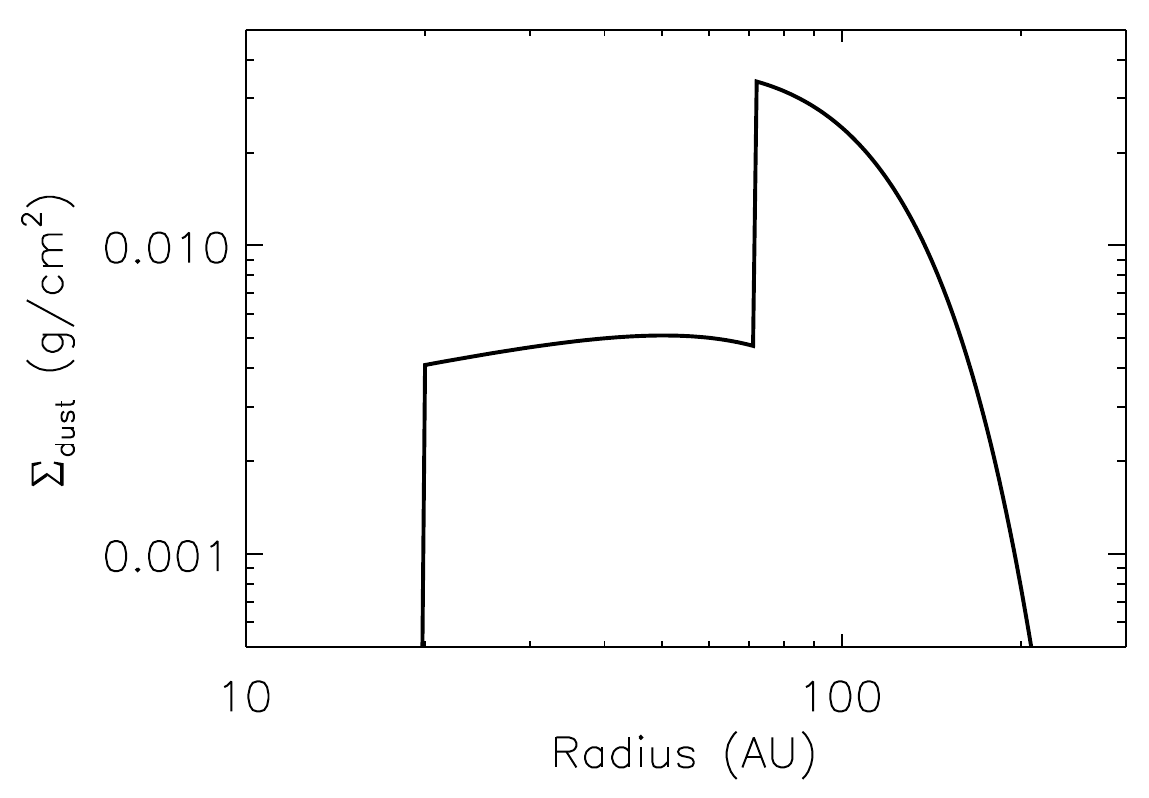}
\caption{We show the surface density function for our adopted disk model.  The surface density within the cavity (20 -- 72 AU) is reduced by a factor $\sim10$ relative to the disk.}
\label{fig:bestsigma}
\end{figure}

In Figure \ref{fig:bestsigma}, we show the dust surface density for our adopted disk parameters.  The total inferred dust mass is $\sim0.1$ $\,M_{\rm Jup}$, in close agreement with the simple calculation presented in \S\ref{sec:mass}.  The inner cavity in this model has a dust mass of $\sim0.007$ $\,M_{\rm Jup}$.

\newpage
\section{Discussion}
\label{sec:discussion}

\subsection{Disk mass}

Our continuum observations and modeling allow us to determine the dust mass in the disk of J1604-2130, and the high disk luminosity rules out the possibility of this being a debris disk.  In addition, the detections of the CO J=3$\rightarrow$2 and J=1$\rightarrow$0 lines indicate the presence of gas in the disk.  However, CO attains high optical depth at low column densities and provides only a lower limit to the gas mass.  A more accurate determination requires multi-line spectroscopy and detailed modeling \citep[e.g.][]{Kamp:2011} and is the subject of a future paper.  To provide a rough estimate for the disk gas mass, we select from the DENT grid of disk models \citep{2010MNRAS.405L..26W} all young (age $\le$10 Myr), solar mass stars with face on 0.1 $\,M_{\rm Jup}$ dust disks and large inner holes.  Among these, the detection of CO J=3$\rightarrow$2 at a level of $\sim10^{-19}$ W/m$^2$ suggests a gas mass of 3 $\,M_{\rm Jup}$ with a factor of 10 uncertainty.

\subsection{Inner disk}
Our adopted model indicates the inner 20--72 AU of the J1604-2130 disk has $0.007$ $\,M_{\rm Jup}$ ~of dust in grains up to $\sim$1 cm in size.  This represents a decrease in surface density of a factor of $\sim$10 relative to the dust rich outer disk shown in our SMA image.  Examination of the parameter space indicates that all feasible models achieve a similar low inner disk dust density, either through this abrupt decrease in surface density or a combination of greater value of the characteristic radius and a much steeper decline in surface density  ($R_C$ as high as $\sim115$ AU and $\gamma$ as low as $\sim -1.5$, respectively).  However, even this extreme requires a factor of 2.5 decrease in surface density at $R_{cav}$.  The dust surface density decreases towards the center of the disk, and the transition cannot be smooth.

The additional truncation of dust near 20 AU is indicated by the rising SED at 16 and 24 $\mu$m and is consistent with IRS spectroscopy which shows no evidence for the 10 or 20 $\mu$m silicate features \citep{2009AJ....137.4024D}, suggesting a deficiency of micron sized grains at $\sim$1 AU scales.  However, there are indications that the inner region is not entirely devoid of material.  Near-infrared spectroscopy from \cite{Dahm:2010} reveals excess emission corresponding to blackbody emission at a temperature of $\sim$900 K.  From \cite{Wyatt:2008}, 
 
 \begin{equation}
 r_{dust}(AU) \approx \left(\frac{278.3}{T}\right)^2 \left(\frac{L}{\Lsun}\right)^{0.5}, 
 \end{equation}
this corresponds to blackbody dust orbiting at approximately 0.08 AU.  \cite{2009AJ....137.4024D} noted a factor of 4 variation between fluxes measured at 4.5 and 8 $\mu$m with Spitzer IRAC, and measurements at those wavelengths a few weeks later with IRS spectroscopy;  these wavelengths roughly correspond to blackbody temperatures between $\sim$1100 and 600 K, which could in turn arise from dust at radii from $\sim$0.05 to 0.17 AU.  

The CO imaging presented here does not rule out the presence of gas within the inner disk region.  Indeed, the inverse P-Cygni profile and 10\% full-width of 252 km/s seen in \cite{Dahm:2011} suggest the system could be accreting, albeit at a very low rate.  \cite{2009AJ....137.4024D} and \cite{Dahm:2011} report H$\alpha$ equivalent widths of -0.57\AA ~and -0.27\AA, respectively; this is too low for the system to be clearly identified as accreting by the spectral type dependent criteria of \cite{White:2003} (3\AA ~for K0 -- K5 stars).  The system similarly shows no emission in the K-band Br$\gamma$ line \citep{Dahm:2010}, another accretion indicator.  Using the relations between accretion and H$\alpha$ luminosity of \cite{Dahm:2008}, we estimate an upper limit on the accretion rate of 10$^{-11}$ \Msun/yr.

\newpage
\subsection{Hole formation mechanism}

Several mechanisms could lead to the formation of an inner hole in a circumstellar disk, summarized in \cite{Williams:2011}.  Recent work has provided a step-by-step classification scheme to identify the dominant mechanism leading to the formation of a transition disk.  \cite{2010ApJ...712..925C} studied a large sample of young transition disks and developed a systematic classification system.  Below, we discuss these potential mechanisms and address their feasibility in explaining the formation of the J1604-2130 transition disk.

\textbf{Grain growth:}  The mm slope of $\alpha_{mm}$=3.17 between 880 $\mu$m and 2.6 mm indicates that substantial grain growth has occurred in the outer disk.  In the warm inner regions of the disk, however, grain growth and dust settling are expected to lead to disks with a smooth downward slope to their infrared excess \citep[e.g.][]{2010ApJ...712..925C}, in contrast to the essentially non-existent excess shown by J1604-2130 at wavelengths up to 16 $\mu m$, where the SED abruptly rises.  

In addition, accretion is expected to continue in disks where an opacity hole has formed due to grain growth.  From the empirical model of \cite{Hartmann:1998}, we would expect a 3 $\,M_{\rm Jup}$ ~disk at 5 Myr to have an accretion rate $\sim3\times10^{-10}$\Msun/yr, though this relation has large scatter for individual objects.  This is more than an order of magnitude higher than the estimated upper limit of $10^{-11}$\Msun/yr for J1604-2130.  While the accretion rate could be consistent with grain growth, the sharp changes in dust surface density suggest that while growth has occurred, it is not likely the primary source of the disk cavity seen in J1604-2130.

\textbf{Photoevaporation:}  As a disk depletes its gas content, the accretion rate is expected to drop as well.  When the accretion rate is comparable to the photoevaporation rate (typically $\sim10^{-10}$\Msun/yr), photoevaporation is expected to become the dominant evolution mechanism for the disk, halt accretion,  and clear the disk from the inside-out.  This is expected to take place when disks have a total mass of 0.05--0.5 $\,M_{\rm Jup}$ ~\citep{Alexander:2006}.  While this is approximately an order of magnitude less than the mass of the J1604-2130 disk, the large uncertainty in the disk mass prevents us from ruling out photoevaporation as the driver of hole formation in this disk.  However, it is unclear that photoevaporation can produce inner regions of depleted but non-zero dust mass.  Such a region is not found in current models of disk structure under the influence of photoevaporation.  Moreover, while the observed H$\alpha$ equivalent width is too low for the system to be formally described as accreting, the inverse P-Cygni profile seen in H$\alpha$ emission \citep{Dahm:2011} suggests that accretion is continuing, albeit at too low a level to be precisely determined.

\textbf{Binarity:}  Disks can form holes due to tidal truncation by a binary companion \citep[e.g. LkCa 4,][]{2008ApJ...678L..59I}).  An equal mass companion opens a disk gap with a radius approximately twice the semi-major axis of the binary system \citep{Artymowicz:1994}.  However, at 36 AU the presence of a binary companion to J1604-2130 has been ruled out to an upper limit of $\sim$20 $\,M_{\rm Jup}$ ~\citep{2008ApJ...679..762K}.  Combining the imaging and aperture-masking interferometry surveys of \cite{2008ApJ...679..762K} and \cite{Ireland:2011}, a companion is ruled out at all radii within the disk down to $\sim$13--20 $\,M_{\rm Jup}$, except for 2 AU and less ($<60$ $\,M_{\rm Jup}$), and $\sim$60--75 AU ($<70$ $\,M_{\rm Jup}$).  However, the limit on companion mass in the $\sim$60--75 AU range is likely much lower than reported, as J1604-2130 would not have exhibited closure phases consistent with a point source in \cite{2008ApJ...679..762K}, and would have triggered further investigation (A. Kraus, personal communication).

\textbf{Planet formation:}  \cite{2010ApJ...712..925C} use the lack of a near-infrared excess, followed by a steeply rising excess, as an indicator for dynamical clearing.  The SED of J1604-2130, which rises steeply from 16 to 24 $\mu m$, matches this criterion.  However, in their study all systems identified as potentially hosting a planet were still actively accreting.  Given that a Jupiter mass planet can intercept up to 90\% of the gas accreting from the outer disk \citep{Lubow:2006}, a more massive planet, or several planets, could possibly be reducing the accretion rate onto the star.  

A single planet will directly clear only a narrow gap in a disk.  However, after gap formation, inner material will accrete onto the star and leave a large gap.  A typical accretion disk model with an evolution time of 1 Myr at 30 AU experiences significant evolution on a timescale of $\sim5$ Myr at 70 AU \citep{Hartmann:1998, Armitage:2011}, suggesting that a single giant planet could have opened the hole in the J1604-2130 disk.  Disk clearing efficiency increases with additional giant planets \citep{2011ApJ...738..131D}.

We have ruled out binarity as the mechanism for the formation of the gap in the J1604-2130 disk, and consider grain growth an unlikely candidate due to the two-region dust surface density function.  The complicated dust distribution, in conjunction with marginal evidence for accretion, makes photoevaporation an unlikely cause for the inner hole.  In our further discussion, we explore possible scenarios for a nascent planetary system.

We can use the upper limit on accretion to estimate a limit to the size of a single giant planet in this system.  Models explore the possibility of a planet truncating the disk, while allowing sparse material to move past \citep[e.g.][]{2007MNRAS.375..500A}.  In such a scenario, the putative planet would be located close to the disk truncation radius; for illustration, we examine the case of a planet located just inside $R_{cav}$, at an orbital radius of (65 AU).  The final truncation of the disk appears to occur at a smaller orbital radius, $R_{gap}$, an issue we discuss below.  

\cite{2007MNRAS.375..500A} derived an expression relating the stellar accretion rate of a protoplanetary system to the disk mass and planetary properties.  We solve their Equation 29 to estimate the mass of a putative planet, assuming the stellar accretion rate is approximately 1/4 the planetary accretion rate:

\begin{equation}
  M_P \simeq 0.25 \frac{M_{d}}{\dot{M}_{acc}} \frac{3\alpha_{acc}\Omega H^2}{2R(R_{out}-R_{in})} M_{Jup},
\end{equation}
with disk mass M$_d$, accretion rate $\dot{M_{acc}}$, accretion parameter $\alpha_{acc}$, orbital frequency $\Omega$, scale height at the planetary orbit $H$, planet orbital radius $R$, and disk inner and outer radii $R_{in}$ and $R{_{out}}$.  From our disk model, we adopt  $R_{in}$ and $R{_{out}}$ of 72 and 200 AU, respectively.  The dimensionless accretion parameter $\alpha_{acc}$ can reasonably have values ranging from 0.001 to 0.01.  

From the H$\alpha$ equivalent width of -0.57 \AA, we estimated an upper limit to stellar accretion of $10^{-11}$\Msun/yr.  At an orbital radius of 65 AU around a 1\Msun ~star, the orbital frequency will be about 0.002 $yr^{-1}$, and assuming the 100:1 gas-to-dust mass ratio of the ISM, the disk has a mass  $M_d = 10^{-2}$ \Msun.  Our adopted disk model gives an estimated scale height, $H$, at 65 AU of less than 3 AU.

Given these assumptions, for an accretion constant $\alpha_{acc}$ = 0.01, a planet at 65 AU with a mass of 6.4 $\,M_{\rm Jup}$ could allow accretion onto the central star at the upper limit rate of $10^{-11}$\Msun/yr.  With an increasing flaring constant, $\psi$, and hence smaller scale height, the necessary planet mass decreases, reaching 6 $\,M_{\rm Jup}$ for $\psi = 1.30$.  A lower value of $\psi$ would lead to a correspondingly higher planet mass, e.g. about 8 $\,M_{\rm Jup}$ for $\psi = 1.05$.  

By comparison, a lower accretion rate could suggest a more massive planet; however, the observation of H$\alpha$ emission in a P-Cygni profile suggests that while stellar accretion may be low, it is unlikely to be significantly lower than this upper limit.  This in turn suggests the mass of such a planet could not be more than a factor of a few larger than 8$\,M_{\rm Jup}$, else accretion would be so low as to eliminate even the hint of  H$\alpha$ emission seen in this system.  In addition, the planet mass scales with $\alpha_{acc}$.  For $\alpha_{acc} = 0.001$, the planet mass could be an order of magnitude lower.

We have so far ignored the truncation of the sparse inner disk at 20 AU.  At this radius, \cite{2008ApJ...679..762K} place an upper limit of 13 $\,M_{\rm Jup}$ on the mass of a companion.  However, our disk model suggests a low dust density, which could be diverted by a low mass planet.  This planet would be in addition to the mechanism causing the decreased dust surface density within 72 AU.

The complex dust structure seen in the J1604-2130 disk --- variable dust at a fraction of an AU, a fully depleted region up to 20 AU, and a partially depleted region from 20 to 72 AU -- may be best explained by the presence of multiple planets.  \cite{Zhu:2011} present a scenario in which four planets lead to a similar configuration, though they must invoke an additional dust filtering mechanism.  \cite{2011ApJ...738..131D} explore similar multi-planet scenarios, where low infrared opacities are driven by geometric filling effects as accretion continues through dynamically shaped streamers.  In both scenarios, multiple planets can achieve low continuum opacities across many tens of AU,  while also reducing stellar accretion more efficiently than single planets.  If planets are diverting accreting material, transition disks such as J1604-2130 may be promising targets for searches for young planets; the additional accretion luminosity will improve the prospects for detection, as in the case of the planet recently imaged within the transition disk of LkCa 15 \citep{2011arXiv1110.3808K}.

\subsection{Comparison to other transition disks}
Surveys with the Spitzer Space Telescope have shown that
$\sim 10-20$\% of circumstellar disks in nearby star forming regions
have mid-infrared SED decrements indicative of an inner hole.
Most of these objects have low millimeter fluxes and are either debris
disks or in the late stages of uv-photoevaporation \citep{Cieza:2008}.
About 20\% have properties that are consistent with ongoing
giant planet formation (Cieza et al. 2012), as we suspect to be the case
with J1604-2130.  Relative to the full protoplanetary disk population,
therefore, such objects are rare.

High resolution millimeter interferometry of these potential planet
forming transition disks has resolved the inner holes in about a dozen
cases to date.  Their masses range from
$0.006 - 0.12\,M_\odot$ and inner radii vary between $15 - 73$\,AU
with no apparent correlation \citep{Andrews:2011c, 2009ApJ...704..496B}.
J1604-2130 lies below the median in terms of mass
but has one of the largest cavities yet found.
Further, it is one of a handful that are known to exhibit strong
line emission in the outer disk \citep[e.g., ][]{2009ApJ...698..131H}.
In a future paper, we will present the rich
chemistry that we have found here, most likely due to the
release of previously depleted species from grain surfaces
at the exposed inner rim \citep{Cleeves:2011}.
It also stands out as having negligible stellar accretion,
which may indicate massive and/or multiple giant planets,
and which may also maintain the large outer disk reservoir over
the relatively advanced, $\sim 5$\,Myr, age of the system.

 \section{Summary}
\label{sec:summary}

In this paper, we have presented high resolution SMA data of the
$880\,\mu$m continuum and CO 3--2 line emission of the transition
disk around J1604-2130, a solar mass star in the Upper Scorpius moving
group.  The images show a nearly face-on ring of dust and gas and we
have determined the dust properties through a least squares fit to
the infrared-millimeter SED and submillimeter visibilities.

We derive a dust mass of $0.1\,M_{\rm Jup}$.  
The millimeter SED slope is a power law out to the longest observed
wavelength of 2.6 mm indicating the power law distribution of
grain sizes extends beyond about 0.8 cm.  The combined visibility-SED
fitting is consistent with these values and determines the inner edge of
the submillimeter dust ring to be 72\,AU in radius.  
The stellar photosphere is seen out to about $10\,\mu$m in wavelength
showing that the inner 20\,AU around the star is dust free,
although variability at shorter infrared wavelengths suggests some
dust may sporadically pass through this region.
Mid-infrared excesses indicate dust emission from the
intermediate zone 20--72\,AU although the radially averaged
surface density is about an order of magnitude lower than the
outer regions.

The presence of millimeter CO emission shows that the outer disk is more primordial
in nature than debris.  The gas-to-dust ratio is at least unity and
probably substantially higher.  Both the CO image and radially averaged
visibilities suggest that the gas may extend slightly closer to the star
than the millimeter emitting dust.

The relatively high mass of the gas-dust disk, its sharp inner edge,
and the lack of gas accretion onto the star together point to a
dynamical origin for the inner hole.  No companions have been found
down to brown dwarf masses and we suggest that at least one and
perhaps several giant planets reside within about 40\,AU radius.

J1604-2130 is the most massive disk remaining in Upper Scorpius
and therefore perhaps the last holdover from the epoch of giant
planet formation.  It may well be that the particular architecture
of its planetary system has conspired to trap a substantial
reservoir of gas and dust at large radii, thereby enabling
this case study of the late stages of primordial disk evolution.

\bigskip
\bigskip
\begin{acknowledgments}
The authors would like to thank the referee and scientific editor for helpful comments.  G.S.M. would like to thank Star Mathews for many thoughtful discussions over the course of writing this paper.  G.S.M. and J.P.W. acknowledge NASA/JPL and NSF for funding support through grants RSA-1369686 and AST08-08144 respectively.  F.M. acknowledge PNPS, CNES and ANR (contract ANR-07-BLAN-0221 and ANR-2010-JCJC-0504-01) for financial support.  
\end{acknowledgments}

{\it Facilities:} \facility{SMA, IRAM:PdBI}

\end{document}